# On the Collapse of Neutron Stars

Jose N. Pecina-Cruz
Intelligent Systems, Inc.
1501 Camellia Ave., McAllen, TX 78501
E-mail: jpecina@Intelligent-e-Systems.com

**Abstract**

This paper reviews the Oppenheimer, Volkoff and Snyder's claim upon the formation of black holes from the collapse of Neutron Stars. It is found that such collapse is inconsistent with microscopic causality and Heisenberg uncertainty principle.

**Introduction**

It has been suggested by Oppenheimer, Volkoff [1] and Snyder [2] that after the thermonuclear sources of energy are exhausted, a sufficient heavy star will collapse. Tolman's gravitational solutions [3], and the Fermi state equation for a gas of cold neutrons [4] are the mathematical framework for this claim. This paper recreates these calculations and finds that microscopic causality and Heisenberg uncertainty principle are inconsistent with a gravitational collapse. It is proposed that an alternate interpretation may exist different to the creation of a black hole. Section 1 is devoted to recreate Einstein's Field Equation for a perfected spherical fluid, section 2 recreates the Fermi State Equation, section 3 recreates equations for a gravitational collapse of a neutron star, and section 4 is devoted to present a new alternative to that of Black Hole formation.

**1. Relativistic Field Equations**

Starting with a static line element exhibiting spherical symmetry of a perfect fluid, Tolman derived the following Einstein's field equations with a null cosmological constant [3].

$$8\pi p = e^{-\lambda(r)}\left(\frac{\nu'(r)}{r} + \frac{1}{r^2}\right) - \frac{1}{r^2}, \tag{1}$$

$$8\pi\rho = e^{-\lambda(r)}\left(\frac{\lambda'(r)}{r} - \frac{1}{r^2}\right) + \frac{1}{r^2}, \tag{2}$$

$$\frac{dP}{dr} = \frac{P + \rho(P)}{2}\nu'(r), \tag{3}$$

Where P is the pressure of a perfect fluid, and $\rho(P)$ is the Fermi state equation.
Using the Schwarzschild's solution for empty space and integrating the last equations by evaluating across the boundary between matter and empty space one obtains

$$\frac{du}{dr} = 4\pi\rho(P)r^2, \qquad (4)$$

where the new variable, $u$, is given by $u = \frac{1}{2}r\left(1 - e^{-\lambda(r)}\right)$

$$\frac{dP}{dr} = -\frac{P + \rho(P)}{r(r - 2u)}[4\pi P r^3 + u]. \qquad (5)$$

## 2. Fermi State Equation for a Degenerate Gas of Cold Neutrons

Let's consider a neutron gas at an absolute temperature of T = $0^0$K. The number of quantum states of the translational motion of a particle with absolute value of its linear momentum between $p$ and $p+dp$. In our case a neutron (fermion), is given by [4].

$$N = g\frac{4\pi V \hat{p}^2 d\hat{p}}{(2\pi\hbar)^3}. \qquad (6)$$

For a gas of neutrons g = 2. The state equation, $\rho(P)$, for a relativist neutron gas completely degenerate (fermion gas at zero absolute temperature) is obtained by

$$E = N\varepsilon = \frac{Vc}{\pi^2\hbar^3}\int_0^{\hat{p}} \hat{p}^2\sqrt{m^2c^2 + \hat{p}^2}\, d\hat{p}, \qquad (7)$$

Integrating, one obtains [4]

$$E = \frac{cV}{8\pi^2\hbar^3}\left\{\hat{p}(2\hat{p}^2 - m^2c^2)\sqrt{\hat{p}^2 + m^2c^2} - (mc)^4 \operatorname{arcsinh}(\frac{\hat{p}}{mc})\right\}$$

Where the energy of a neutron, $\varepsilon = \sqrt{m^2c^4 + \hat{p}^2c^2}$, was replaced in the above equation. For the pressure one obtains [4]

$$P = -\left(\frac{\partial E}{\partial V}\right)_{s=0}, \qquad (8)$$

$$P = \frac{c}{8\pi^2\hbar^3}\left\{\hat{p}(\frac{2}{3}\hat{p}^2 - m^2c^2)\sqrt{\hat{p}^2 + m^2c^2} + (mc)^4 \operatorname{arcsinh}(\frac{\hat{p}}{mc})\right\}, \qquad (9)$$

A parametric form of the equation of state [4] is

$$\rho = \frac{E}{V} = K(\sinh t - t), \tag{10}$$

$$P = \frac{1}{3}K(\sinh t - 8\sinh\frac{1}{2}t + 3t), \tag{11}$$

With

$$K = \frac{\pi m^4 c^5}{4\hbar^3}, \tag{12}$$

And

$$t = 4\log\left(\frac{\hat{p}}{mc} + \left[1 + \left(\frac{\hat{p}}{mc}\right)^2\right]^{\frac{1}{2}}\right), \tag{13}$$

This Fermi's state equation, ρ(P), is constrained by microscopic causality. According to microscopic causality, local densities D(x) of observable operators do not interfere and therefore commute for space-like separations [5]

$$[D(x_1), D(x_2)] = 0, \qquad (x_1 - x_2)^2 < 0, \tag{15}$$

Microscopic causality is sizeable only for distances comparable to the Compton wavelength of the particle. Since the information can not propagate faster than the speed of light. That is,

$$\lambda = \frac{\hbar}{mc}. \tag{16}$$

$\lambda \sim 10^{-15} m$ for a neutron. Therefore a relativistic neutron gas must satisfy this constraint. The Fermi state equation can not be valid for neutrons separated by a distance shorter than their Compton wavelength. Beyond this distance a new change phase must occur to a collapsing neutron star. This alternative to black hole formation is suggested in this manuscript.

### 3. Fermi's Relativistic State Equation for a Collapsing Neutron Star

Substituting P and ρ(P) into Eqs. (4) and (5) and changing to relativistic or Plank units, $c = G = \hbar = 1$. One obtains the state equation for a perfect fluid with spherical symmetry

$$\frac{du}{dr} = r^2(\sinh t - t) \tag{17}$$

And

$$\frac{dt}{dr} = -\frac{4}{r(r-2u)} \frac{\sinh t - 2\sinh\frac{1}{2}t}{\cosh t - 4\cosh\frac{1}{2}t + 3} \quad (18)$$

$$\times \left[\frac{1}{3}r^3(\sinh t - 8\sinh\frac{1}{2}t + 3t) + u\right]$$

These equations are to be integrated from $u = u_0, t = t_0$ at $r = 0$ to $r = r_b$ where $t_b = 0$, and $u = u_b$. Authors of Reference 1, found five numerical solutions, for five values of $t_0$, of the relativistic field equations (17) and (18). These equations have been obtained by substituting Fermi's state equation $\rho$ and $P$ into the Einstein's field equation for a perfect fluid with spherical symmetry. Authors of Ref 1 also found that for a cold neutron core there are not static solutions, for core masses greater than a mass $M_{max} \sim 0.76\odot$ and a minimum radius of $R_{min} = 9.42 km$. Ref 1 discusses the final behavior of very massive neutron stars reaching the following conclusion: "<u>either the equation of state we have used so far fails to describe the behavior of highly condensed matter that the conclusions are qualitative misleading</u>[1] or the star will continue to contract indefinitely, never reaching equilibrium." They disregard the first option [2] and conclude that a gravitational collapse occurs or a black hole formation.

### 4. An Alternative to Black Holes Formation

Because equations (17) and (18) are constrained by microscopic causality a gravitational collapse is not permissible. That is, the neutrons in the relativistic Fermi gas must be separated by a distance greater than their Compton wavelength.
An alternative to a gravitational collapse is the neutron-antineutron pair formation. The uncertainty principle for a relativistic point particle [6] is given by

$$\Delta x = \frac{\hbar}{mc}, \quad (19)$$

Therefore when one tries to assign coordinates (the center of the star R=0) to a falling particle (neutron) in a collapsing star one needs to provide to this particle a momentum of the same order than its rest mass, thus allowing the creation of a new particle (particle-antiparticle) [7]. Also if one tried to find out the behavior of two neutrons, of the relativistic Fermi gas, separated by a distance equal to its Compton wavelength, one needs to provide to one of the neutrons a energy on the order of $mc^2$, allowing again the creation of particles (pairs). This simple second argument shows that in a relativistic quantum mechanics, the coordinates of a particle can not appear as dynamic variables.
According to the arguments above the final fate of a neutron star is the neutron-antineutron pair creation and posterior disintegration. That is, starburst of pions and gamma rays must be emitted by a "collapsing" neutron star. Some possible modes of disintegration could be

$$n + \bar{n} -> p + \bar{p},$$

$$p + \bar{p} -> 4(\pi^+ + \pi^-), \quad (20)$$

$$n + \bar{n} -> \gamma^{(-)} + \gamma^{(+)}.$$

---

[1] Underlined by the author

**Conclusion**

This paper suggests a new interpretation of the final fate of a neutron star. Although a clearer description must rest on a quantum theory of gravity.

**Acknowledgment**

This manuscript is dedicated to Cris Villarreal for her encouragement and her conviction that this paper could have a repercussion in physics. I also dedicate this work to Roger Maxim Pecina on his 18th birthday.